# Phase shift and magnetic anisotropy induced field splitting of impurity states in $(Li_{1-x}Fe_x)OHFeSe$ superconductor

Tianzhen Zhang[1†], Yining Hu[1†], Wei Su[2,7†], Chen Chen[1†], Xu Wang[1], Dong Li[3], Zouyouwei Lu[3,10], Wentao Yang[1], Qingle Zhang[1], Xiaoli Dong[3,10,11], Rui Wang[4,8*], Xiaoqun Wang[5], Donglai Feng[6,1,8,9], Tong Zhang[1,8,9*]

[1] Department of Physics, State Key Laboratory of Surface Physics and Advanced Material Laboratory, Fudan University, Shanghai 200438, China
[2] Beijing Computational Science Research Center, Beijing 100084, China
[3] Beijing National Laboratory for Condensed Matter Physics, Institute of Physics, Chinese Academy of Sciences, Beijing 100190, China
[4] National Laboratory of Solid State Microstructures and Department of Physics, Nanjing University, Nanjing 210093, China
[5] School of Physics, Zhejiang University, Hangzhou 310058 Zhejiang, China
[6] National Synchrotron Radiation Laboratory and Department of Physics, University of Science and Technology of China, Hefei 230026, China
[7] College of Physics and Electronic Engineering, Center for Computational Sciences, Sichuan Normal University, Chengdu 610068, China
[8] Collaborative Innovation Center for Advanced Microstructures, Nanjing 210093, China
[9] Shanghai Research Center for Quantum Sciences, Shanghai 201315, China
[10] School of Physical Sciences, University of Chinese Academy of Sciences, Beijing 100049, China
[11] Songshan Lake Materials Laboratory, Dongguan, Guangdong 523808, China

**Revealing the energy and spatial characteristics of impurity induced states in superconductors is essential for understanding their mechanism and fabricating new quantum state by manipulating impurities. Here by using high-resolution scanning tunneling microscopy/spectroscopy, we investigated the spatial distribution and magnetic field response of the impurity states in $(Li_{1-x}Fe_x)OHFeSe$. We detected two pairs of strong in-gap states on the "dumbbell" shaped defects. They display clear damped oscillations with different phase shifts and a direct phase – energy correlation. These features have long been predicted for classical Yu-Shiba-Rusinov (YSR) state, which are demonstrated here with unprecedented resolution for the first time. Moreover, upon applying magnetic field, all the in-gap state peaks remarkably split into two rather than shift, and the splitting strength is field orientation dependent. Via detailed numerical model calculations, we found such anisotropic splitting behavior can be naturally induced by a high-spin impurity coupled to anisotropic environment, highlighting how magnetic anisotropy affects the behavior of YSR states.**

The response of superconductivity to impurities relies on the nature of electron pairing and the properties of impurity. An impurity with magnetic moment has exchange coupling with conducting electrons, which generally induces pair-breaking scattering and give rise to the Yu-Shiba-Rusinov (YSR) states [1-3]; meanwhile, nonmagnetic impurities could induce in-gap states in unconventional superconductors [4]. Historically, impurity states have played important role in exploring the order parameter and quasi-particle band structure of superconductors [4-8], and they also have important applications such as constructing topological nontrivial superconductivity and Majorana quasi-particles through inter-impurity coupling [9-13].

Scanning tunneling microscope (STM) can directly measure the localized impurity states via tunneling spectrum (dI/dV). These states appear as one or multiple pairs of peaks inside superconducting gap situated symmetrically around $E_F$ [4,14-25]. Besides, STM is capable of measuring spatial distributions of the impurity state. As shown by Rusinov [3], YSR state wavefunction should exhibit (damped) $k_F$ oscillation with a phase shift directly corelated to the binding energy. Such oscillation plays a critical role in the coupling between impurities [12], and carries information on the Fermi surface of the host. So far, although oscillated YSR state has been detected in few STM studies [18,21], a quantitative study on their properties (especially about the phase shift), is still limited.

Despite similar appearance of in-gap peaks, the specific formation mechanism of impurity states could be diverse. For magnetic impurities, YSR states originate from electron transits between the ground state and the lowest excited spin state. The ground state depends on the strength of exchange coupling versus superconducting pairing [4,16,24-26], and can be further affected by magnetocrystalline anisotropy [19]. Meanwhile, anisotropic scattering potential can generate multiple YSR states with different angular momenta, which complicates the interpretation [4,15]. Since impurities or adsorbates on the surface usually have low symmetric environment and variable exchange couplings strength, their ground state could be different and induce YSR states with different spin/orbital properties [15-16,21-24]. Therefore, more diagnostic measurement is desired to identify the origin of impurity states, which is important for understanding their superconducting host and designing novel quantum state based on them.

As predicted by theory, probing the response of impurity states to external magnetic fields can help to clarify their origin [27]. Depending on the degeneracy of the ground/excited state and the selection rules, YSR state peaks either shift or split under magnetic field [27]. However, such measurement would require high energy resolution because Zeeman energy is usually small (≈0.1meV at $B$=1T) and the superconductivity must persist under relatively high field. So far, clear field responses of YSR states are only observed in $K_xFe_2Se_2$ film [17].

In this work, we studied the native impurity (Fe vacancy) induced states in an iron-based superconductor $(Li_{1-x}Fe_x)OHFeSe$ by using low-temperature STM. We obtained their spatial distribution and response to magnetic field with unprecedented details. The two pairs of observed impurity states show clear oscillations which obey the YSR wavefunctions of a quasi-2D system, and display a direct YSR phase–energy correlation. It is remarkable that all the impurity state peaks split into two upon applying magnetic field, and the splitting strength is field orientation dependent. Via detailed modeling and numerical calculations, we show that the observed behaviors can be naturally induced by an unscreened high-spin ($S$=2) in the presence of magnetic anisotropy.

The experiment was performed in a dilution refrigerator STM (Unisoku) at an effective electron temperature of 160 mK [28]. High quality $(Li_{0.8}Fe_{0.2})OHFeSe$ single crystalline films were grown on $LaAlO_3$ substrate by matrix-assisted hydrothermal epitaxy [29]. The sample was cleaved in ultra-high vacuum and normal Pt/Ir tips were used after being treated on Au(111) surface.

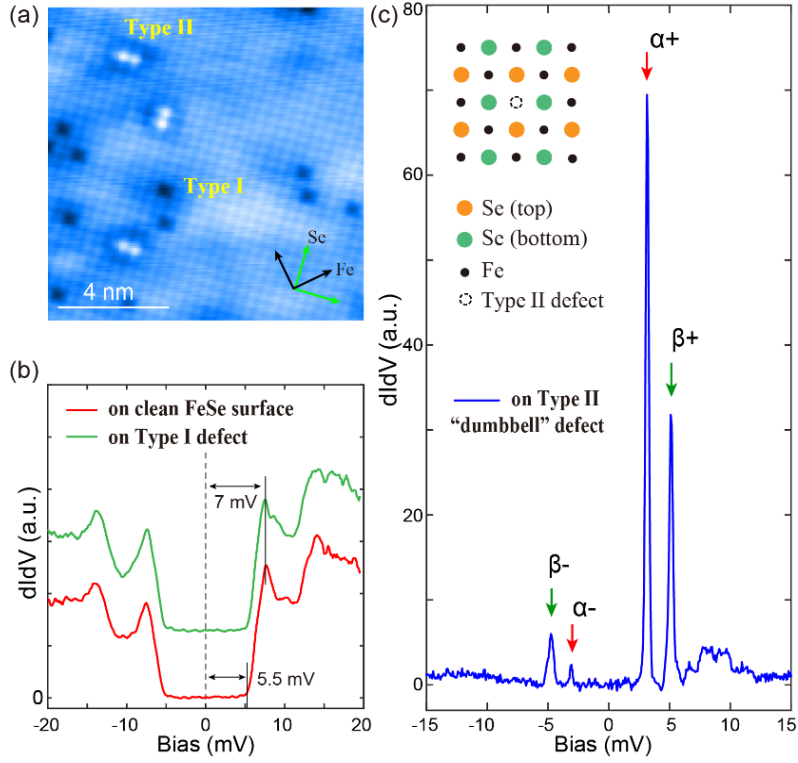

**FIG. 1** (a) Topographic image of the FeSe surface of (Li,Fe)OHFeSe ($V_b$=160 meV, I=10 pA). (b) Superconducting gap spectra taken on defect-free region and Type I defect (Se vacancy). (c) Normalized dI/dV spectra taken on a dumbbell-shaped Type II defect ($V_b$ = 15 meV, I=60 pA). Two pairs of sharp in-gap states are observed (inset: atomic structure of dumbbell defect).

Figure 1(a) shows a topographic image of FeSe-terminated surface of $(Li_{0.8}Fe_{0.2})OHFeSe$. There are two types of commonly observed defects: the Se vacancy (Type I) and the "dumbbell" shaped defects centered at Fe sites (Type II). On defect free region, dI/dV spectrum shows a full superconducting gap with two coherence peaks at ±7 and ±14 meV [Fig. 1(b)]. The flat region of gap bottom has a half width of 5.5 meV, which corresponds to the minimum gap value. We find Se vacancies do not induce any in-gap states [Fig. 1(b)], while two pairs of strong in-gap states (with asymmetric intensities) are observed at dumbbell defect site with energies of ±3.1 meV and ±5.0 meV [Fig. 1(c)]. Below we refer them as α± and β± states, respectively. We note the dominant defects in hydrothermally synthesized (Li,Fe)OHFeSe which can suppress superconductivity are reported to be Fe vacancies [30]. Therefore, the dumbbell defects are expected to be mostly Fe vacancies.

The Fe vacancy has been shown to carry magnetic moment in $K_xFe_2Se_2$ [17], whose electronic structure is similar to that of (Li,Fe)OHFeSe. Fig. 1(c) inset sketches the atomic lattice near a Fe vacancy. Notably, the local environment of such Fe-site defect is only two-fold symmetric. Low symmetric environment is expected to introduce anisotropic scattering potential and magnetocrystalline anisotropy for a local spin.

Then dI/dV mapping was preformed around a single dumbbell defect to reveal local density-of-states (LDOS) distribution. Figs. 2(a,b) and (c,d) show the dI/dV maps measured at energies of α±, β± states, respectively [see Fig.S1 of supplementary materials (SM, ref. 31) for additional data]. Clear oscillation pattern is observed and persists up to ~10 nm away from the defect, reflecting the LDOS variation of impurity states. The oscillations generally show higher intensity along Fe-Fe directions. This is due to the Fermi surface anisotropy of (Li,Fe)OHFeSe and will be discussed later. Figs. 2(e,f) show the dI/dV line profiles of α± and β± along Fe-Fe direction, respectively. One can immediately see damped oscillations with a period of ≈1.9 nm for both α± and β±, while the oscillation phase varies for different states. Notably, the phase difference between α+, α- is larger than that between β+, β-, and the decay of α± is faster than β±.

To have quantitative analysis, we simulate the LDOS with classical YSR wavefunction [4]:

$$\Psi_\pm(r) \propto \frac{\sin(k_F r + \delta_\pm)}{r^{\frac{d-1}{2}}} \exp\left[-|\sin(\delta_+ - \delta_-)|\frac{r}{\xi}\right]. \quad (1)$$

Here $r$ is the distance to defect, $k_F$ is Fermi wavevector, $\xi$ is superconducting coherence length, $\delta_\pm$ are scattering phase shift of positive and negative energy states, and $d$ is the dimensionality of the system. The LDOS is then given by $|\Psi(r)|^2$ with a $2k_F$ oscillation and $2\delta$ phase. By taking $k_F$=0.165 Å$^{-1}$, $\xi$=3.0 nm, $d = 2$, $\delta_{\alpha+} - \delta_{\alpha-}$=0.30π, $\delta_{\beta+} - \delta_{\beta-}$=0.15π, the simulated LDOS can well fit the line profiles for $r \gtrsim 1$ nm in Figs.2(g,h). The faster decay of α± is naturally accounted by their larger phase difference, which is a characteristic of quasi-particle bound state. Meanwhile $d = 2$ indicates a 2D like electron band, which is consistent with the quasi-2D structure of (Li,Fe)OHFeSe [29,30]. The deviation of simulated LDOS at $r \lesssim 1$ nm could be due to that $Eq.$ 1 is an asymptotic solution, which may not apply for small $r$.

Furthermore, the scattering phase difference of a pair of YSR states is predicted to directly related with their energy via:

$$|\varepsilon_l| = \Delta_0 \cos(\delta_{l+} - \delta_{l-}). \quad (2)$$

Here $l$ stands for different scattering channels. We find the measured energies of α±, β± and their phase difference well satisfy this relation. A corresponding fitting in Fig. 2(g) yields $\Delta_0$ = 5.51 meV (which is close to the minimal gap value found in Fig.1(b)), and the fitting error of $\varepsilon$ is smaller than ±0.1 meV. Therefore, both the energy and spatial properties of α±, β± match the characteristics of YSR state. Their similar oscillations period suggest they are from the same band. Since dumbbell defect is an anisotropic scatterer, α±, β± are likely from two scattering channels with different quantum numbers. This will be described in the following.

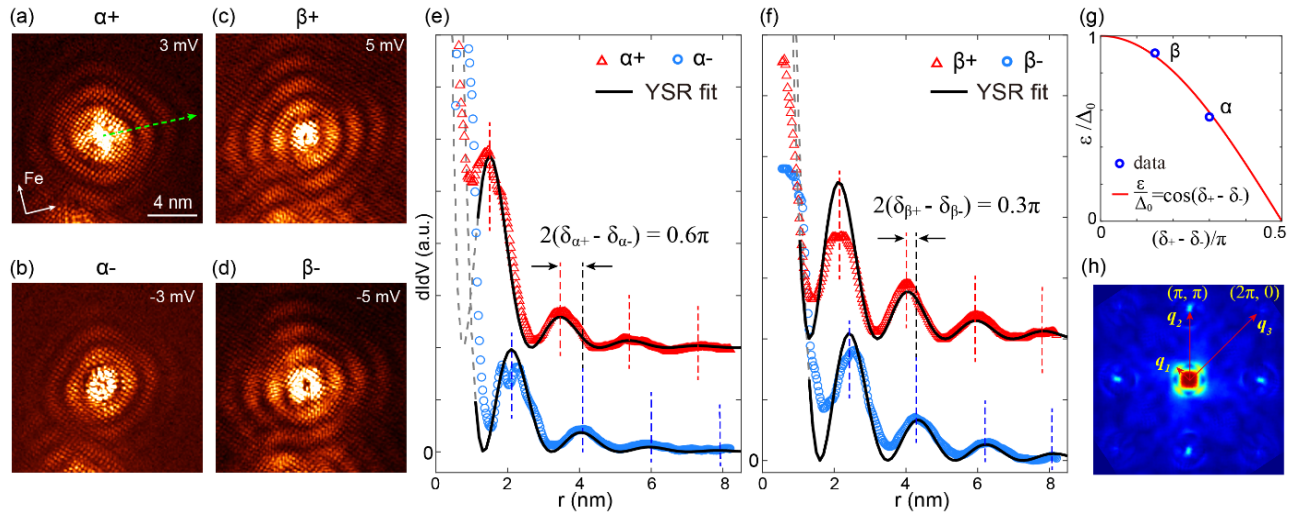

**FIG. 2** (**a-d**) dI/dV maps around a dumbbell defect, taken at energies of α±, β± states. (**e, f**) dI/dV line profiles of α± and β± along Fe-Fe direction (green arrow in (a)), respectively. The YSR LDOS fittings are shown by black solid curves. (**g**) The energies of α±, β± as function of their phase difference, and a fitting by using Eq.2. (**h**) The FFT of (a).

The spatial extension of impurity state can also be treated as quasi-particle interference generated by impurity, which carries information on the Fermi surface structures [7,8]. Fig. 2(h) shows the fast-Fourier transform (FFT) of α+ state. It is similar to the reported normal QPI pattern of (Li,Fe)OHFeSe [32,33]. The $2k_F$ oscillation gives rise to the "ring" structure near (0,0) with enhanced weight along Fe-Fe. As discussed in ref. [32,34], such feature originates from inter-pocket scattering of the oval-shaped electron pockets (see Fig.S2 and Part I of SM [31] for more details). The value of $k_F$ measured here (0.165 Å$^{-1}$) agrees with the ARPES measurement [35,36]. Besides, the ring-like feature near (π, π), (2π, 0) and equivalent points are also present. They are from the scattering between electron pockets.

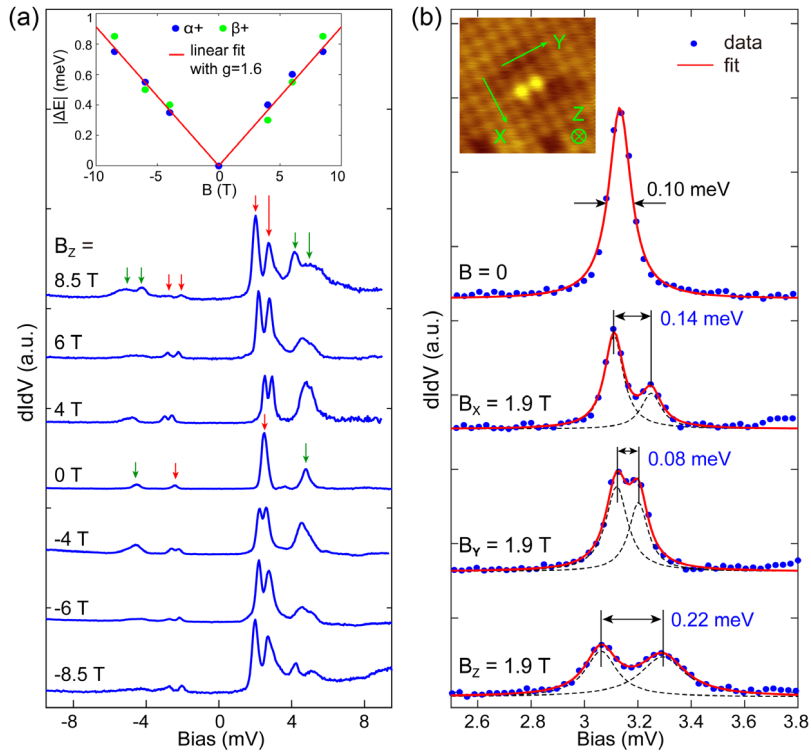

**FIG. 3.** (**a**) Tunneling spectra taken on dumbbell defect under various out-of-plane ($B_Z$) fields. The splitting of α± and β± peaks is indicated by arrows. Inset shows the splitting magnitude of α+ and β+ versus field strength. (**b**) Tunneling spectra taken under $B$ along different directions (indicated in inset), with energy range around α+. Red solid curves are Lorentzian peak fittings.

Next we investigated the response of the impurity states to magnetic field. Fig. 3(a) shows the dI/dV spectra of a dumbbell defect taken under out-of-plane field ($B_Z$), with the strength from $B_Z$ = -8.5 T to +8.5 T. Remarkably, all the in-gap state peaks split into two in the field (indicated by arrows, more clearly seen for α±). This is quite different from the shifting YSR peaks observed in K$_x$Fe$_2$Se$_2$ (110) film [17]. At certain |$B_Z$|, the splitting magnitude (|Δ$E$|) is similar for α and β and for opposite $B_Z$ directions (Fig. 3(a) inset). Assuming the splitting is

Zeeman like ($\Delta E = g\mu_B B$), a linear fitting yields a Landé factor of $g = 1.6$. We note that at high fields, the two split peaks have different intensity (the high energy one has low intensity). This feature is addressed by our model calculation below.

We further examined the field orientation dependence of the splitting. Fig. 3(b) shows a set of spectra taken under magnetic fields with the same strength (1.9T) but along three orthogonal directions of $B_X$, $B_Y$ and $B_Z$ (see inset image for their definition). A spectrum taken at B=0 is also shown for comparison. The energy range of these spectra is centered around α+ state with optimized resolution (individual peak width is typically of 0.10 meV). One can clearly see the splitting is different for different orientation: the out-of-plane field ($B_Z$) gives the largest $\Delta E$ of ~0.22 meV, while an in-plane field parallel to the dumbbell ($B_Y$) gives the smallest $\Delta E$ of ~0.08 meV.

To understand the behaviour of the YSR states, we performed model calculations by using density-matrix renormalization group (DMRG). The impurity system is described by the Anderson model, $\mathcal{H} = \mathcal{H}_{SC} + \mathcal{H}_{imp} + \mathcal{H}_{hyb}$, where $\mathcal{H}_{SC}$ is the BCS mean-field Hamiltonian. The hybridization between the conduction electrons and the impurity is given by $\mathcal{H}_{hyb} = \sum_{\boldsymbol{k},m,\sigma}(V_{\boldsymbol{k}} c_{\boldsymbol{k},\sigma}^\dagger d_{m,\sigma} + \text{h.c.})$, where $d_{m,\sigma}$ is the annihilation operator for the impurity state with magnetic quantum number $m$ and spin $\sigma$. $\mathcal{H}_{imp}$ describes the local impurity state with the on-site repulsion [37].

The dumbbell defect has twofold symmetry; the coupling $V_{\boldsymbol{k}}$ satisifes $V_{\boldsymbol{k}} = V_{k,\phi} = V_{k,\phi+\pi}$, where $\phi$ is the angle of the 2D wavevector $\boldsymbol{k}$. After making expansion of the conduction electron states to the spherical harmonics, this symmetry would impose restrictions to $m$ in $\mathcal{H}_{hyb}$. For demonstration, we take $V_{k,\phi} = V_k \cos^2\phi$ and make Schrieffer-Wolff transformation to $\mathcal{H}$. A multi-channel s-d exchange model is obtained as $\mathcal{H}_{sd} = \sum_{k,k',\sigma,\sigma',m} J_m c_{k,m,\sigma}^\dagger \boldsymbol{\sigma}_{\sigma\sigma'} \cdot \boldsymbol{S}_{imp} c_{k',m,\sigma'}$, where $\boldsymbol{S}_{imp}$ is the impurity spin operator, and the sum of the channel $m$ only involves $m = 0, \pm 2$. Interestingly, the $m = \pm 2$ channels are coupled to each other in $\mathcal{H}_{SC}$, which are not directly coupled to $m = 0$. Thus, two different energy scales for the in-gap states would emerge, which could account for the observation of the α± and β± states. In addition, the field orientation dependent splitting in Fig.3(b) is reminiscent of magnetic anisotropy effect, such as what observed in spin excitations and Kondo resonance of surface magnetic atoms [38,39]. Hence, we take into account additional terms, $\mathcal{H}_{ani} = \mathcal{D}S_{imp,z}^2 + \mathcal{E}(S_{imp,y}^2 - S_{imp,x}^2) + g\mu_B \boldsymbol{B} \cdot \boldsymbol{S}_{imp}$, where $\mathcal{D}$ and $\mathcal{E}$ terms account for the axial and transverse anisotropies respectively, and $g\mu_B \boldsymbol{B} \cdot \boldsymbol{S}_{imp}$ describes the Zeeman effect [31].

We first consider the single-channel case with only $m = 0$. Since $S_{imp}$ is hard to measure directly, we consider different values of $S_{imp}$ from 1/2 to 5/2. The formation of YSR states and their response to magnetic fields can be analyzed qualitatively. It is known that the competition between exchange coupling and superconducting pairing leads to a quantum phase transition at $J = J_c$ [4]. When $J < J_c$, an unscreened phase is generally found; while for $J > J_c$, a Kondo screened phase appears. In the unscreened (screened) phase, the impurity ground state is an $S_{imp}(S'_{imp})$ multiplet where $S'_{imp} = S_{imp} - 1/2$; while its first excited state is an $S'_{imp}(S_{imp})$ multiplet. At the microscopic level, YSR states originate from the processes where an electron transits from the ground to the excited states and vice versa, with the spin selection rule of

$\Delta S_z = \pm 1/2$. Taking an $S_{imp} = 1/2$ system for example [Figs. 4(a,b)], in the unscreened phase, the ground state is a doublet and the excited state is singlet ($S'_{imp} = 0$). Magnetic field can split the ground state but only the $S_z = -1/2$ to $0$ transition is allowed, so the YSR state only shifts under the field [Fig. 4(a)]. In the screened phase, the transitions from $S_z = 0$ to $\pm 1/2$ are both allowed, so the YSR state will split under magnetic field [Fig. 4(b)].

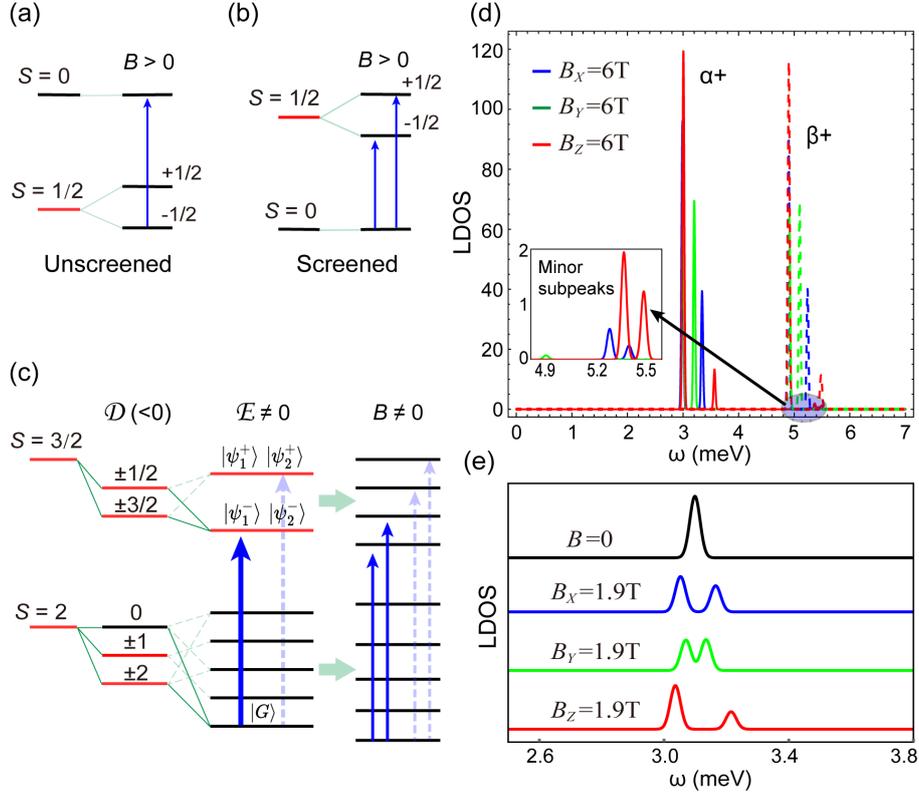

**FIG. 4.** (**a,b**) Schematics of ground/excited states of $S=1/2$ impurity in the unscreened and screened phase (respectively), and their responses to magnetic field B. (**c**) Schematics of $S=2$ impurity in the unscreened phase for the single-channel model. The effects of anisotropy $\mathcal{D}$, $\mathcal{E}$ and $B$ are shown. The blue arrows denote the dominant processes that contribute to the in-gap LDOS. Light arrows are the higher-order processes, which are negligible (see panel d). (**d**) The full LDOS obtained by DMRG for the multi-channel model with $S=2$, showing both α+ and β+ state under finite $B_Z$, $B_X$ or $B_Y$. Zoom-in data of shaded region displays the minor subpeaks originated from higher-order processes in (c). (**e**) Calculated LDOS shows the anisotropic splitting under $B_Z$, $B_X$, $B_Y$=1.9T.

For higher spin systems ($S_{imp} > 1/2$), magnetic anisotropy can further lift the degeneracy of spin-multiplets, which endows the YSR state with more intricate splitting features. Here we take $S_{imp} = 2$ for example (Fig. 4(c)). In the unscreened phase and $\mathcal{D}<0$, the ground state splits into $S_z = \pm 2, \pm 1, 0$ ($S_z = \pm 2$ has the lowest energy) and the excited state split into $S_z = \pm 3/2, \pm 1/2$. With a nonzero $\mathcal{E}$ term, the spin states are further hybridized. The states $S_z = \pm 2$ hybridize with $S_z = 0$, generating the ground state $|G\rangle$. The excited states also hybridize and form two pairs of doubly degenerate states, $|\psi_1^-\rangle, |\psi_2^-\rangle$ and $|\psi_1^+\rangle, |\psi_2^+\rangle$. The in-gap YSR state is mainly contributed by the transition process between $|G\rangle$ and $|\psi_1^-\rangle, |\psi_2^-\rangle$, as indicated by the thick blue arrow in Fig.4(c). Under field $B$, $|\psi_1^-\rangle$ and $|\psi_2^-\rangle$ further split, inducing two in-gap YSR states corresponding to the two transitions marked by thin blue arrows.

Interestingly, both $\mathcal{D}$ and $\mathcal{E}$ terms play important roles. Their combination not only enables the splitting under $B$, but also breaks the spin rotation symmetry and lead to different splitting strengths for $B$ along different directions, which explains our experimental observation. A complete analysis for different $S_{imp}$ is shown in Fig. S5. Generally, we find that with nonzero $\mathcal{D}$ and $\mathcal{E}$, anisotropic splitting occurs either for an integer $S_{imp}$ in the unscreened region or for a half-integer $S_{imp}$ in the screened region. However, experimentally we did not observe signatures of the screening (Kondo resonance) at the temperatures above $T_C$ (see Fig. S3), therefore a screened phase is disfavored here.

Finally, we take all channels $m$ into account, and numerically solve the multi-channel s-d exchange models. We employ DMRG to calculate the ground state as well as a few excited states. As shown in the LDOS (for $\omega > 0$) in Fig.4(d), two pairs of in-gap states corresponding to the α+ and β+ states are found, Interestingly, these states split with different strength under $B_X$, $B_Y$ and $B_Z$. Notably, in the shaded region of Fig.4(d) and the inset, minor subpeaks are also found, which split under external fields. These subpeaks, which originate from the higher-order processes indicated by the light arrows in Fig.4(c), are negligible compared to the α+ and β+ states. More splitting details of the α+ state are shown in Fig. 4(e). The anisotropic behavior is clearly observed, and the split is largest for $B_Z$ and then for $B_X$ and $B_Y$. Besides, both the peak positions and the splitting values are in quantitative agreement with that in Fig.3(b). Furthermore, a higher peak is generally found at lower energy after splitting, in consistent with the spectra of Fig. 3(a,b). More details are presented in Part II of SM [31], which show quantitative agreement between the simulation and experiments.

In summary, by combining the measurement of LDOS distribution, magnetic field response and model calculations, we show the in-gap states induced by dumbbell defect in (Li,Fe)OHFeSe can be attributed to YSR state of a local spin with significant magnetic anisotropy. Our study therefore demonstrates a practical way to reveal the nature of impurity induced in-gap state, which have diverse origins as shown in our model. Our direct measurement of a scattering phase-energy relation and field induced splitting also deepens the understanding on classical YSR state.

Note added: Upon preparing this manuscript, we noticed a recent high-resolution STM study on YSR state also observed field induced splitting [40].

†These authors contributed equally.
*Corresponding authors. Email: rwang89@nju.edu.cn, tzhang18@fudan.edu.cn

**Acknowledgments:**
We thank professors Qiang-Hua Wang, Xi Dai and Zhongxian Zhao for helpful discussion. This work is supported by the Innovation Program for Quantum Science and Technology (Grant no.: 2021ZD0302800), National Key R&D Program of the MOST of China (Grant no.: 2022YFA1402701), National Natural Science Foundation of China (Grants Nos.: 92065202, 1222540, 11888101, 11961160717, 11974244, 11904245, 12274206, 12104094, 12061131005,

11834016), Science Challenge Project (Grant No. TZ2016004), Shanghai Municipal Science and Technology Major Project (Grant No. 2019SHZDZX01), Shanghai Pilot Program for Basic Research, Fudan University 21TQ1400100 (21TQ005), the Strategic Priority Research Program of Chinese Academy of Sciences (Grants No. XDB33010200 and XDB25000000), and Xiaomi Foundation.

# Supplementary Materials of "Phase shift and magnetic anisotropy induced field splitting of impurity states in (Li$_{1-x}$Fe$_x$)OHFeSe superconductor"

Tianzhen Zhang†, Yining Hu†, Wei Su†, Chen Chen†, Xu Wang, Dong Li, Zouyouwei Lu, Wentao Yang, Qingle Zhang, Xiaoli Dong, Rui Wang*, Xiaoqun Wang, Donglai Feng, Tong Zhang*

†These authors contributed equally.
*Corresponding authors. Email: rwang89@nju.edu.cn, tzhang18@fudan.edu.cn

## Part I: Additional data on the dumbbell defect induced in-gap states

Figure S1(b~h) show additional dI/dV maps around the single dumbbell shaped defect taken at various energies [Fig. S1(a) is the topographic image]. Only at the energies of the in-gap state peaks the LDOS modulation is significant. Fig. S1(i) shows the evolution of dI/dV spectra as moving away from defect center. The coherence peaks are significantly suppressed at the defect site while start to show up at ~0.6 nm away from the defect.

Figure S2(a,b) show the FFT image of the LDOS map of α+, β+ states, respectively. They display similar features to the normal state QPI observed in (Li,Fe)OHFeSe in ref. 31 [as the one shown in Fig. S2(d)]. There is ring-like feature near (0,0) with high intensity along Γ-M direction. Such features are from inter-band scattering of anisotropy electron pockets. As sketched in Fig. S2(c), the bulk Fermi surface of (Li,Fe)OHFeSe composes of oval-shaped electron pockets at M points (which is similar to that of 1ML FeSe/SrTiO$_3$). Due to the Brillouin zone *folding,* each M point should have two electron pockets; however the folded pocket has much weaker spectral weight [*e.g.*, as reported in PRL 117, 117001 (2016)] and the band hybridization between the two pockets has not been directly observed. The electron pocket has multiple orbital components: the long axis portion is contributed by $d_{xy}$ orbital (blue) and the short axis parts are from $d_{yz}/d_{xz}$ orbitals (red/green). In normal state QPI, the intra- and inter-pockets scatterings (marked by $q_1 \sim q_3$) can be observed as ring like structure near (0, 0), (π, π) and (2π, 0). In an STM measurement, the tunneling probability of $d_{yz}/d_{xz}$ orbitals is higher than that of the planar $d_{xy}$ orbital. This is evidenced by the "incomplete" ellipse near (2π, 0) in Fig. S2(d) whose long axis portion is missing. We note that similar QPI pattern was also observed in single-layer FeSe/SrTiO$_3$ [*e.g.*, Nat. Phys. 11, 946 (2015), Phys. Rev. Lett. 115, 017002 (2015)] and (Li,Fe)OHFeSe bulk crystals [ref.32], which has been carefully addressed by a multi-orbital simulation [Phys. Rev. B 93, 125129 (2016)].

It is worth to mention that the ARPES study in PRL 117, 117001 (2016) observed two local gap maxima around the single electron pocket in 1ML FeSe/STO. Such *k*-space gap anisotropy can also result two coherence peaks (corresponding to two gap maxima) in the DOS. Thus it is possible that the double-peaked gap of (Li,Fe)OHFeSe is from (or mainly from) a single electron pocket. This is in line with the observation that α and β states appear to have a single band origin.

Figure S3 shows the dI/dV spectra taken at the dumbbell defect site and defect-free region, measured at T= 78K (above the Tc of (Li,Fe)OHFeSe). The defect spectra show some difference but no obvious Kondo resonance peak is observed near $E_F$.

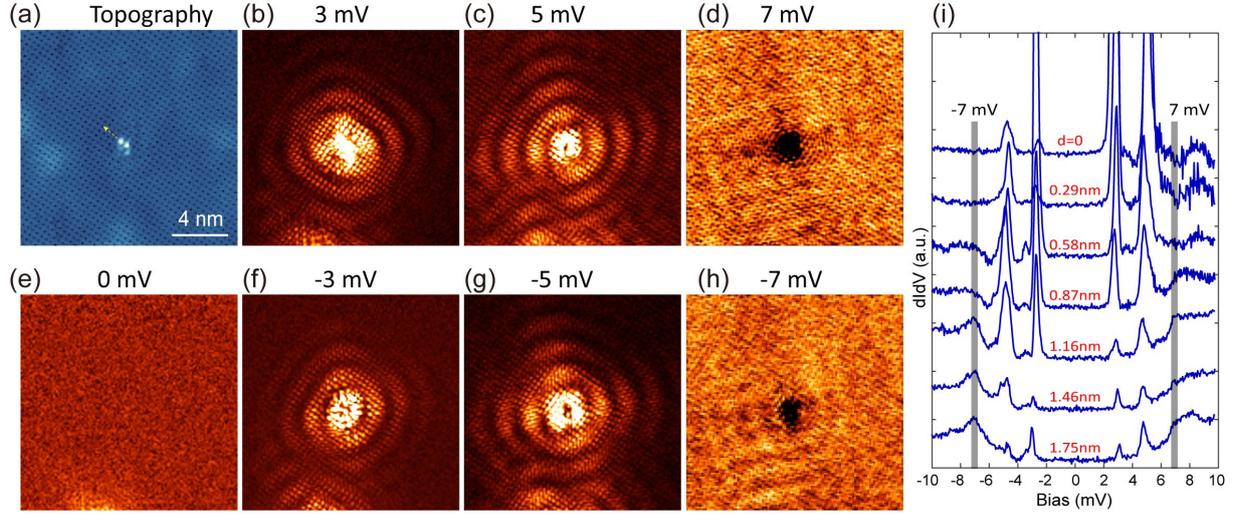

**Fig. S1.** (a) Topographic image around a single dumbbell defect (scan size: 16nm x 16nm). (b~h) dI/dV maps taken at different energies, as labelled above each panel (set point: $V_b$ = 6 mV, I = 100 pA). (i) A series of dI/dV spectra taken on a line from the center of dumbbell defect (along the arrow in panel (a)), with 0.29 nm steps. The two vertical bars indicate the position of coherence peaks at ±7meV.

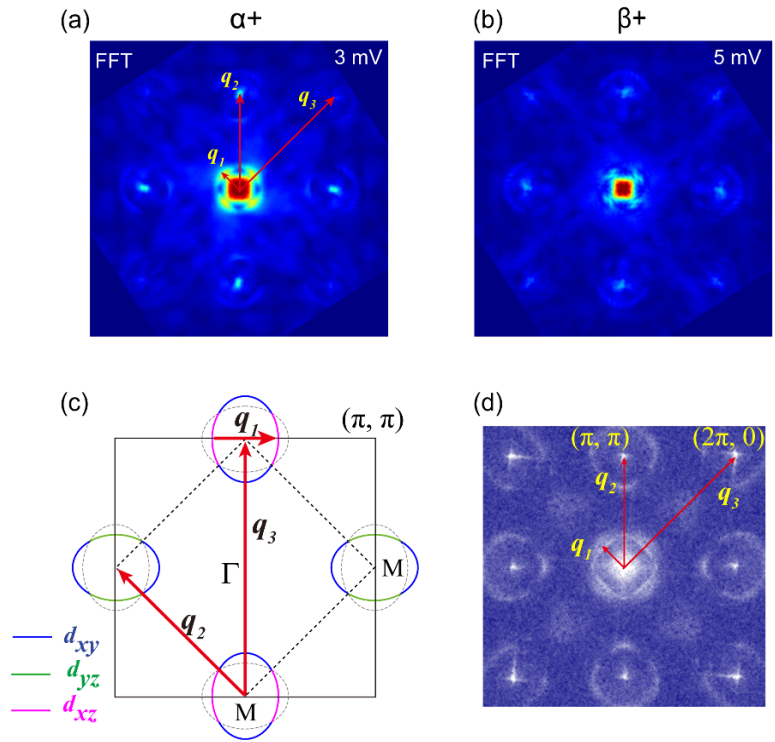

**Fig. S2.** (a, b) Four-fold symmetrized FFT of the LDOS map of α+, β+ states, respectively. (c) The bulk Fermi surface sketch of (Li, Fe)OHFeSe, different orbital components are indicated by different colors. (d) Normal state QPI taken at E= 20meV (adapted from ref. [31]).

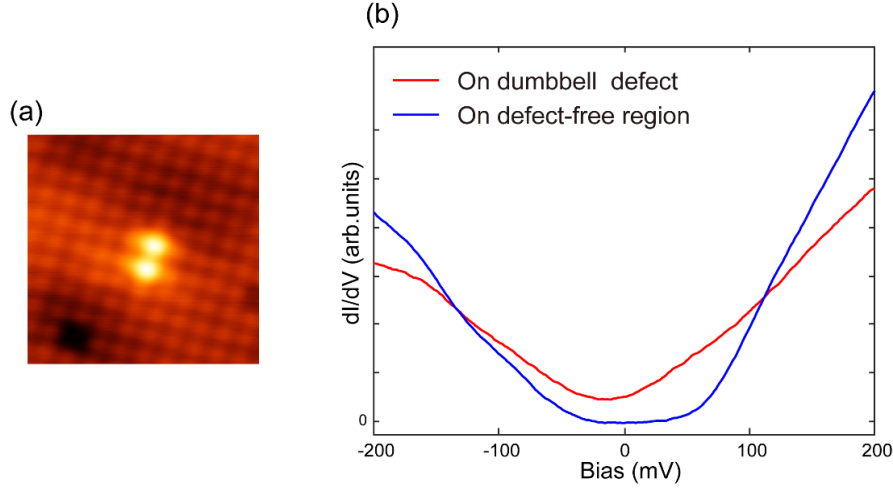

**Fig. S3.** (a) Topographic image of a dumbbell defect. (b) dI/dV spectra taken at the defect site (red) and defect-free region (bule), measured at T= 78K (above the Tc of (Li,Fe)OHFeSe). The defect spectra show some difference but no obvious Kondo resonance peak is observed near $E_F$.

**Part II: Theoretical analysis on the Magnetic anisotropy and Zeeman splitting effects and numerical renormalization group results.**

**1. Method and analysis of the Single-Channel model:**

To focus on the effect of magnetic anisotropy, we firstly consider in this section a single channel model that captures the main physics, where the impurity is coupled to the conduction electrons of a superconductor with a single channel *m*. A more realistic case taking into the multi-channel components will be derived and studied in details in the following sections.

The impurity system is described by the Hamiltonian $\mathcal{H} = \mathcal{H}_{SC} + \mathcal{H}_{imp} + \mathcal{H}_B$. $\mathcal{H}_{SC}$ describes the mean-field BCS Hamiltonian of superconductor (SC):

$$\mathcal{H}_{SC} = \sum_{\mathbf{k},\sigma} \varepsilon_{\mathbf{k}} c^\dagger_{\mathbf{k},\sigma} c_{\mathbf{k},\sigma} + \Delta \sum_{\mathbf{k}} (c^\dagger_{\mathbf{k},\uparrow} c^\dagger_{-\mathbf{k},\downarrow} + h.c.) \tag{1}$$

The s-d coupling between impurity and the SC is given by $H_{imp}$:

$$\mathcal{H}_{imp} = J\mathbf{S}_{imp} \cdot \mathbf{s} + \mathcal{D} S^2_{imp,z} + \mathcal{E}(S^2_{imp,y} - S^2_{imp,x}) \tag{2}$$

Here we consider an impurity with magnetic anisotropy. $\mathbf{S}_{imp}$ is the impurity spin with $S_{imp,\alpha}$ its $\alpha$-th component ($\alpha = x, y, z$). $\mathbf{s}$ is the spin operator of the electrons of the SC bath, which is locally coupled to the impurity via antiferromagnetic spin exchange ($J > 0$). The $\mathcal{D}$ term describes the out-of-plane anisotropy, while $\mathcal{E}$ captures the in-plane anisotropy, leading to an easy axis in the x-y plane. We further take into account the external magnetic field $\mathbf{B}$, which generates Zeeman splitting of impurity spin as:

$$\mathcal{H}_B = g\mu_B \mathbf{B} \cdot \mathbf{S}_{imp} \tag{3}$$

where $\mu_B$ is the Bohr magneton, and $g$ is the Landé g-factor.

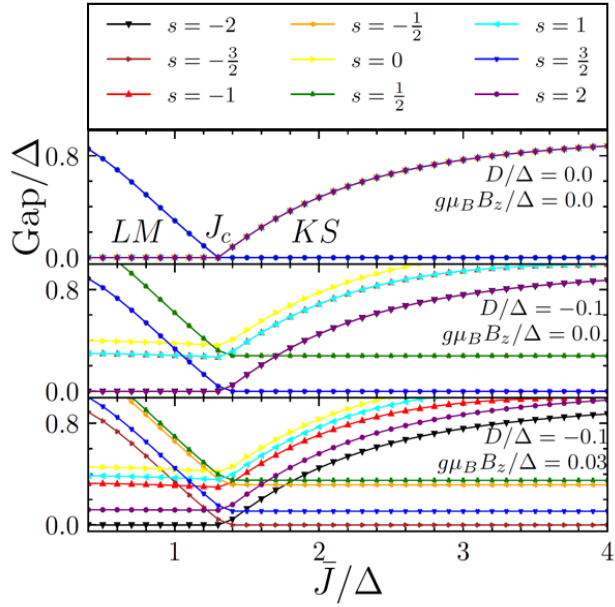

**Fig. S4.** NRG results for the in-gap spectrum of a spin-2 Kondo impurity in SC as a function of the coupling constant $\bar{J} = J\rho$. Here $\rho$ is the density of states at $\varepsilon_F$ for the normal state. The SC bath is logarithmic discretized with parameter $\Lambda = 2$ and we keep ~2048 states in iterative diagonalization with the length of the Wilson chain up to 20. Other parameters are used as $B_x = B_y = 0$, $\mathcal{E} = 0$, and $\Delta = 0.1$.

It is known that there exist non-negligible competitions between the pairing and the spin-exchange coupling for the conduction electrons, which leads to a quantum phase transition at $J = J_c$ between a local momentum (LM) phase and a screened phase. For $J < J_c$, the Cooper pairs dominate over the spin screening, thus the impurity becomes effectively decoupled from the bath in low-energy. In contrast, for $J > J_c$, the local impurity becomes strongly coupled to the bath electrons in low-energy, which locally breaks the Cooper pairs, generating either the fully screened Kondo singlet for $S_{imp} = 1/2$ or the underscreened state for $S_{imp} > 1/2$. The latter effectively displays the remaining spin, $S'_{imp} = S_{imp} - 1/2$. For brevity, we generally term the fully screened and underscreened cases as the screened phase in the following.

The above quantum phase transition between the unscreened (LM phase) and screened phase is manifested by an energy level crossing at $J = J_c$. We use the numerical renormalization group (NRG) to calculate the in-gap spectrum for a spin-2 Kondo impurity in SC. In the NRG calculations, the SC bath is logarithmic discretized with parameter $\Lambda=2$, such that a length of ~10 for the Wilson chain is already long enough to obtain in-gap states with convergence (for the details of the mapping, see Part II-4). We keep ~2048 states in iterative diagonalization, and a Wilson chain with the length of 20 sites is used for the sake of accuracy. Fig. S4 shows the results for the in-gap spectrum with increasing $J$, which clearly displays the transition point $J_c$ for $\mathcal{D} = B_z = 0$. With further turning on $\mathcal{D}$ and $B_z$, we observe splitting of the NRG spectrum. From the calculated NRG spectrum, the local density of states (LDOS) can be readily understood. In the following, we will focus on the behavior of LDOS under magnetic fields with considering the effect of $\mathcal{D}$ and $\mathcal{E}$ for different values of $S_{imp}$ on each side of $J_c$.

At zero temperature, the LDOS in the Lehmann representation can be obtained as:

$$\rho_f(\omega) = \sum_{m,\sigma}[|\langle\varphi_m|f_\sigma^\dagger|\varphi_0\rangle|^2 \delta(\omega + E_0 - E_m) + |\langle\varphi_0|f_\sigma^\dagger|\varphi_m\rangle|^2 \delta(\omega + E_m - E_0)] \quad (4)$$

where $E_0$ ($E_m$) is the energy of the ground state $|\varphi_0\rangle$ (m-th excited state $|\varphi_m\rangle$). The first (second) term creates (annihilates) an electron from the ground state, which contributes to the peaks in $\omega > 0$ ($\omega < 0$) energy range. The positions of the peeks are thus identified to be the energy gaps between the excited states and the ground state.

## 2. Magnetic Anisotropy effects on the splitting of LDOS:

In the unscreened (screened) region, the impurity state can be regarded as a multiplet with spin $S_{imp}$ ($S'$). Correspondingly, the first excited in-gap state is the multiple with spin $S'$ ($S_{imp}$) spin. The observation provides a simple way to analyze the effect of magnetic anisotropy and Zeeman splitting. Using $S_z$ as the quantization axis, the spin multiplets are labeled by its $S_z$ components. Then, we note that the following properties are satisfied by the impurity state with spin $S$:

1) The $\mathcal{D}$ term splits a spin multiplet into the sectors of $S_z = 0, \pm 1/2, \pm 1, \ldots$. For $S > 1/2$, the $SU(2)$ symmetry is no longer conserved in presence of $\mathcal{D}$, only leaving $S_z$ as the conserved quantity. For $S \leq 1/2$, the $\mathcal{D}$ term only leads to a trivial shift of the energy of the spin multiplet.
2) The $B_z$ term splits the spin multiplet into different $S_z$ sectors. Only $S_z$ is conserved for $S \geq 1/2$.
3) The $\mathcal{E}$ term hybridizes the component $S_z = m$ with $S_z = m \pm 2$. It has no effects on an $S \leq 1/2$ spin. For $S > 1/2$, $S_z$ is no longer conserved.
4) The $B_x$ and $B_y$ term hybridize the component $S_z = m$ with $S_z = m \pm 1$. $S_z$ is no longer conserved.
5) For an $S_z$ conserved system, supposing the ground states has $S_z = m$, creation or annihilation of an electron in the ground state only leads to the transition from the ground state to $S_z = m \pm 1/2$. Thus, the transition is between the multiplet of spin-$S'$ and that of spin-$S$. When the $S_z$ conservation is violated by the $\mathcal{E}$ or the $B_x$ or $B_y$ term, higher order transitions also occur, but the transitions between the states within the multiplet of $S$ or $S'$ are still forbidden.

We demonstrate the above effects of $\mathcal{D}$, $\mathcal{E}$, and $B$ by the schematic plot in Fig. S5. Interestingly, significantly different behaviors can be found for the spin-half and spin-integer cases. In general, for the spin-integer case ($S_z = M/2$) with $M$ being an even integer, the $M+1$ fold degeneracies of the multiplet are completely lifted after applying $\mathcal{D}$ and $\mathcal{E}$. Then, with further applying small external magnetic fields $B$, the energy levels would only be shifted, as shown in Fig. S5. However, for the spin-half case, there generally remains a twofold degeneracy for each level after applying $\mathcal{D}$ and $\mathcal{E}$. This is because, for spin-$M/2$ with $M$ being an odd number, the hybridization among the states with $S_z = -M/2, -M/2 + 2, \ldots M/2 - 1$ leads to the bonding/anti-bonding states which have the same energy as those generated by the hybridization among $S_z = -M/2 + 1, -M/2 + 3, \ldots M/2$. As a result, the additional external fields $B$ will further lift the remaining twofold degeneracy, resulting in the energy splitting of the LDOS peaks.

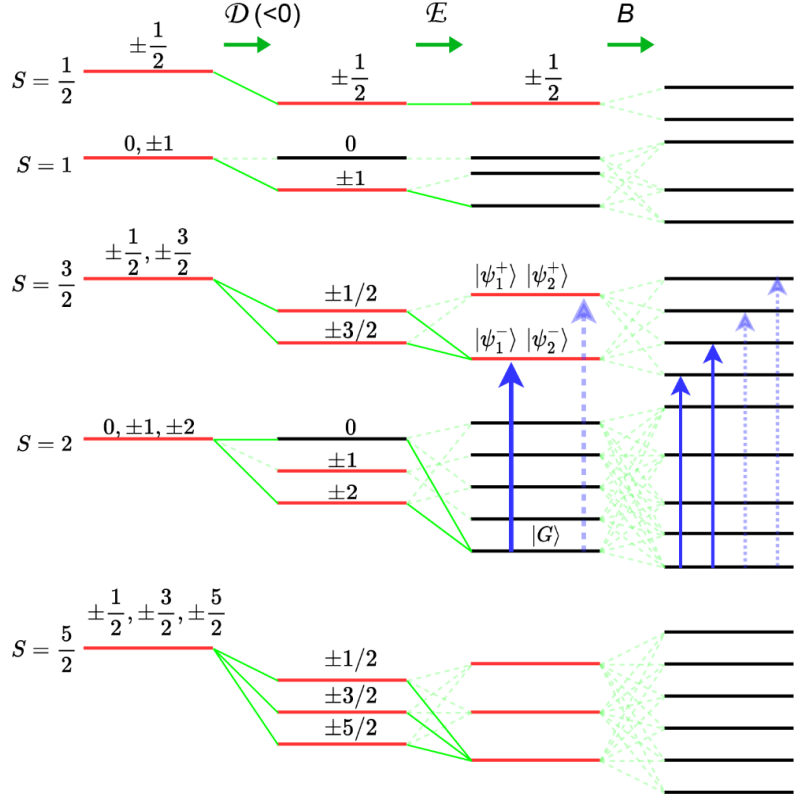

**Fig. S5.** Schematics of the splitting of a spin-$S$ multiplet under the out-of-plane anisotropy $\mathcal{D}$, in-plane anisotropy $\mathcal{E}$ and magnetic field $B$. Red (Black) lines denote states with (without) degeneracy. For a spin-2 impurity in the LM region, the LDOS are contributed by the transition processes between the $S = 2$ and $S = 3/2$ states. The solid blue arrows denote the dominant transition processes at zero temperature, while the dashed blue arrows represent for the high-order contributions negligible in the LDOS. The evolution processes of the lowest spin states under $\mathcal{D}$ and $\mathcal{E}$ are highlighted by the solid green lines.

From the above analysis, one can conclude that the magnetic-field-induced splitting of the LDOS can only take place when the in-gap excited states form a multiplet with spin $M/2$, and $M$ being an odd integer. The situation occurs either in the screened region ($J > J_c$) for the impurity with half-integer spins or in the LM region for the impurity with integer spins. An example is demonstrated by Fig. S5. For a spin-2 impurity in the LM region, the ground state has $S = 2$ and the first excited state has $S = 3/2$. After taking into account $\mathcal{D}$ and $\mathcal{E}$, both the $S = 2$ and $S = 3/2$ multiplets split. The states $S_z=\pm 2$ hybridize with $S_z=0$, generating the ground state $|G\rangle$. The excited states $S_z = \pm 3/2, \pm 1/2$ hybridize and form two pairs of doubly degenerate states, $|\psi_1^-\rangle$, $|\psi_2^-\rangle$ and $|\psi_1^+\rangle$, $|\psi_2^+\rangle$. Correspondingly, two pair of in-gap peaks in LDOS will show up, which reflect the two transition processes marked by the solid and dashed blue arrows in Fig. S5. We note that transition from $|G\rangle$ to $|\psi_1^-\rangle$, $|\psi_2^-\rangle$ has dominant weight compared that from $|G\rangle$ to $|\psi_1^+\rangle$, $|\psi_2^+\rangle$, since the latter is originated from higher-order processes enabled by the $\mathcal{E}$ term. As a result, the in-gap YSR state is mainly contributed by the transition between $|G\rangle$ and $|\psi_1^-\rangle$, $|\psi_2^-\rangle$. Under magnetic field $B$, $|\psi_1^-\rangle$ and $|\psi_2^-\rangle$ further split, giving rise to two in-gap YSR states corresponding to the two transitions marked by the thin blue arrows.

## 3. The Multi-Channel impurity model in superconductor bath:

Based on the above method and the analysis on the single-channel case, we now derive a more realistic impurity model relevant to the experiments, and then present the details in our numerical calculations. In realistic cases, the impurity states of a given orbital angular momentum (OAM) $l$ is composed of $2l + 1$ orbits with different magnetic quantum numbers, $m = -l, -l + 1, ..., l - 1, l$. The total spin $S$ is determined by the Hund's rule as well as how many electrons are filled in the $2l + 1$ states. In the exchange scattering process between the impurity and the conduction electrons, the OAM index $l$ is conventionally regarded as conserved [1]. The microscopic process that plays the key role here describes the hopping of a conduction electron with spin $\sigma$ onto the state $|m, \sigma\rangle$ and then an electron with the opposite spin $|m, -\sigma\rangle$ back to the conduction bath, which generates the spin-exchange between the conduction electrons and the local impurity states. The hopping between conduction electrons and the impurity state $|m, \sigma\rangle$ is naturally described by the Hamiltonian,

$$\mathcal{H}_{hyb} = \sum_{k,m,\sigma}(V_k c^\dagger_{k,\sigma} d_{m,\sigma} + \text{h.c.}), \qquad (5)$$

where $d_{m,\sigma}$ is the annihilation operator for the local impurity state with magnetic quantum number $m$ and spin $\sigma$, and $V_k$ is the hybridization between the conduction electrons and the impurity state. The total Hamiltonian of the impurity system is then given by the Anderson impurity model $\mathcal{H} = \mathcal{H}_{SC} + \mathcal{H}_{imp} + \mathcal{H}_{hyb}$, where $\mathcal{H}_{SC}$ is given by Eq. (1), and $\mathcal{H}_{imp}$ reads as,

$$\mathcal{H}_{imp} = \sum_{m,\sigma} \varepsilon_{d,m} d^\dagger_{m,\sigma} d_{m,\sigma} + U \sum_m n_{d,m,\uparrow} n_{d,m,\downarrow}, \qquad (6)$$

where $\varepsilon_{d,m,\sigma}$ denotes the energy of local orbitals and the Hubbard $U$ describes the cost of energy to add an electron to the filled $m$-th orbital. Then, we will derive the s-d exchange model from the above multi-orbital Anderson model $\mathcal{H}$, with taking advantage of the fact that the impurity scattering has two-fold symmetry evidenced in our experiments.

We firstly represent the conduction electronic states by spherical harmonics. Since our system has a quasi 2D structure, the full spherical harmonic representation can be simplified into $c_{\mathbf{k},\sigma} = c_{k,\phi,\sigma} = \frac{1}{\sqrt{2\pi k}} \sum_m e^{-im\phi} c_{m,\sigma}(k)$, where $\phi$ is the polar angle of $\mathbf{k}$, and $m$ denotes the magnetic quantum number [2]. Then, the two-fold symmetry of the dumbbell impurity should be reflected by the hybridization term $V_{\mathbf{k}} = V_{k,\phi}$, such that $V_{k,\phi} = V_{k,\phi+\pi}$. A simple modeling of $V_{\mathbf{k}}$ is given by $V_{k,\phi} = V_k \cos^2\phi$, which captures the dumbbell geometry of the impurity. Inserting $V_{k,\phi}$ and then making transformation to the spherical harmonics, $\mathcal{H}_{SC}$ can be cast into:

$$\mathcal{H}_{SC} = \sum_{m,k}[\sum_\sigma \varepsilon_k c^\dagger_{m,\sigma}(k) c_{m,\sigma}(k) + \Delta_m c^\dagger_{m,\uparrow}(k) c^\dagger_{-m,\downarrow}(k) + \Delta^*_m c_{-m,\downarrow}(k) c_{m,\uparrow}(k)], \qquad (7)$$

Where $\Delta_m = (-1)^m \Delta$ describes the pairing amplitude between the $-m$ and $m$ channel. Correspondingly, the hybridization term is transformed into:

$$\mathcal{H}_{hyb} = \sum_{k,\sigma,m=0,\pm 2}[V_{k,m}c_{m,\sigma}^{\dagger}(k)d_{m,\sigma} + h.c.], \tag{8}$$

where the conservation of $m$ is implied [1]. $V_{k,m}$ is the Fourier component of $V_k$, and only the $V_{k,m=0}$ and $V_{k,m=\pm 2}$ components are relevant, due to the scattering $V_{k,\phi} = V_k \cos^2\phi$. Therefore, only $m=0$ and $m=\pm 2$ need to be taken into account in Eq. (7), since we are only interested in the impurity physics here.

From Eq. (6)-(8), the Schrieffer-Wolff transformation can be performed, which leads to the s-d exchange model [1], i.e.,

$$\mathcal{H}_{sd} = \sum_{k,k',m} J_m \boldsymbol{\sigma}_{k,k'}^m \cdot \boldsymbol{S}_{imp}, \tag{9}$$

where $J_m$ is the s-d exchange coupling of the m-th channel. $\boldsymbol{\sigma}_{k,k'}^m$ is the spin density operator of the conduction electrons for the m-th channel, namely,

$$\boldsymbol{\sigma}_{k,k'}^m = \sum_{\sigma,\sigma'} c_{m,\sigma}^{\dagger}(k)\,\boldsymbol{\sigma}_{\sigma\sigma'} c_{m,\sigma'}(k') = \sum_{\sigma,\sigma'} A_{0,\sigma}^{m\dagger}\boldsymbol{\sigma}_{\sigma\sigma'} A_{0,\sigma'}^m, \tag{10}$$

where $\boldsymbol{\sigma}_{\sigma\sigma'}$ denotes the Pauli matrix and we introduced $A_{0,\sigma}^m = \sum_k c_{m,\sigma}(k)/\sqrt{N}$ with $\sqrt{N}$ the normalization factor. In addition, the magnetic anisotropy as discussed in Part II-1 should also be taken into account in the impurity Hamiltonian, i.e.,

$$\mathcal{H}_{ani} = \mathcal{D}S_{imp,z}^2 + \mathcal{E}(S_{imp,y}^2 - S_{imp,x}^2) \tag{11}$$

Thus, the total Hamiltonian to be solved is given by $\mathcal{H}_{tot} = \mathcal{H}_{SC} + \mathcal{H}_{sd} + \mathcal{H}_{ani}$, i.e., Eq. (7), Eq. (9) and Eq. (11). Clearly, $\mathcal{H}_{tot}$ is a multi-channel (in terms of magnetic quantum number $m$) Kondo model in superconductor bath with magnetic anisotropy. In the following sections, we will demonstrate that $\mathcal{H}_{tot}$ provides a satisfying quantitative explanation of all our experimental observations.

## 4. Details for numerical calculations of the multichannel Kondo model:

We now use the standard procedure to discretize the superconductor bath [3]. Since the fixed point of the impurity states is determined by low-energy physics near the Fermi level $\varepsilon_F$, we consider the energy band in the energy window $[\varepsilon_F - D, \varepsilon_F + D]$ near $\varepsilon_F$, where $D$ is the energy cutoff which is set to 1 in the calculations. One can make expansion of the energy dispersion of the normal state near $\varepsilon_F$, with keeping the linear term, $v_F(k - k_F)$ and $v_F = k_F/m$. Then, the sum of $k$ is substituted by the integral within the cutoff, i.e., $\int_{-1}^{1} dk$.

Following the conventional mapping approach, we introduce effective fermion operators $f_{n,\sigma}^{m\dagger}$ ($f_{n,\sigma}^m$) defined on the $n$-th site of the Wilson open chain for the $m$-th channel, and $\mathcal{H}_{SC} + \mathcal{H}_{sd}$ can be mapped after tridiagonalization as $\mathcal{H}_{SC} + \mathcal{H}_{sd} = \bar{H}_K + \bar{H}_\Delta + \bar{H}_I$, where

$$\bar{H}_K = \frac{1+\Lambda^{-1}}{2}\sum_{m,\sigma}\sum_{n=1}^{\infty}\Lambda^{-\frac{n}{2}}t_n\left(f_{n,\sigma}^{m\dagger}f_{n+1,\sigma}^m + h.c.\right), \tag{12}$$

$$\bar{H}_I = \sum_{\sigma,\sigma'} \bar{J}_m f_{0,\sigma}^{m\dagger} \boldsymbol{\sigma}_{\sigma\sigma'} f_{0,\sigma'}^m \cdot \boldsymbol{S}_{imp}, \tag{13}$$

$$\bar{H}_\Delta = -\Delta \sum_m \sum_{n=0}^{\infty} (f_{n,\uparrow}^{m\dagger} f_{n,\downarrow}^{-m\dagger} + h.c.), \tag{14}$$

where $\bar{J}_m = J_m \rho$ and $\rho$ the density of states at $\varepsilon_F$ for the normal state, and $\Lambda$ is the NRG discretization factor. The hopping term between the nearest sites on the Wilson chain is obtained as

$$t_n = (1 - \Lambda^{-(n+1)})(1 - \Lambda^{-(2n+1)})^{-1/2}(1 - \Lambda^{-(2n+3)})^{-1/2}. \tag{15}$$

To make further simplification, we apply the Bogoliubov transformation, $b_{n,\uparrow}^{m\dagger} = (f_{n,\uparrow}^{m\dagger} + f_{n,\downarrow}^m)/2$ and $b_{n,\downarrow}^m = (f_{n,\uparrow}^{m\dagger} - f_{n,\downarrow}^m)/2$, where $b_{n\uparrow}^{m\dagger}$ and $b_{n\downarrow}^m$ are the operators for Bogoliubov quasi-particles, and then perform a particle-hole transformation [3], $c_{2n,\uparrow}^{m\dagger} = b_{2n,\uparrow}^{m\dagger}$, $c_{2n,\downarrow}^m = b_{2n\downarrow}^m$ and $c_{2n-1,\uparrow}^{m\dagger} = b_{2n-1,\downarrow}^m$, $c_{2n-1,\downarrow}^m = -b_{2n-1,\uparrow}^{m\dagger}$, which leads to:

$$\bar{H}_K = \frac{1+\Lambda^{-1}}{2} \sum_{m,\sigma} \sum_{n=1}^{\infty} \Lambda^{-\frac{n}{2}} t_n (c_{n,\sigma}^{m\dagger} c_{n+1,\sigma}^m + h.c.), \tag{16}$$

$$\bar{H}_I = \sum_{\sigma,\sigma'} \bar{J}_m c_{0,\sigma}^{m\dagger} \boldsymbol{\sigma}_{\sigma\sigma'} c_{0,\sigma'}^m \cdot \boldsymbol{S}_{imp}, \tag{17}$$

$$\bar{H}_\Delta = -\Delta \sum_{n=0}^{\infty} (-1)^n (\sum_{m,\sigma} c_{n,\sigma}^{m\dagger} c_{n,\sigma}^{-m} - 1), \tag{18}$$

where the sum over $m$ only involves $m = 0$ and $m = \pm 2$, which is clear from the discussions below Eq.(8). Eq.(16)-(18) constitute the mapped chain model to be numerically solved, which involves different channels in terms of the magnetic quantum number $m$. Notably, the $m = 2$ and $m = -2$ channel are coupled to each other via the $\bar{H}_\Delta$ term in Eq. (18), and $\bar{J}_{m=2} = \bar{J}_{m=-2} = \bar{J}_2$ is implied by the two-fold symmetry. Thus, effectively, there are two channels $m = 0$ and $m = \pm 2$ that are not directly coupled, and each of them displays a characteristic energy scale for which Kondo physics would emerge. This feature will be responsible for the emergence of two pairs of LDOS peaks, as observed in experiments (the $\alpha$, $\beta$ states in the main text). Moreover, based on our analysis in Sec.2, the S=2 impurity could provide a most direct explanation of the experiments. Thus, we will focus on the S=2 case in the following.

Then, we solve the mapped model Eq.(16)-(17) together with $\mathcal{H}_{ani}$ in Eq.(11) using DMRG. $U(1) \times U(1)$ symmetry corresponding to the conservation of total particle number for the $m = 0$ chain and the $m = \pm 2$ chains is exploited to reduce the calculation complexity. Different from what we did in Sec. II.1, we use density matrix renormalization group (DMRG) instead of NRG here, since the former has much higher efficiency if only the ground state and a few in-gap states are to be evaluated. As will be clear below, the high efficiency is of key importance here because we need to extract the LDOS with continuously varying the model parameters, ($\mathcal{D}$, $\mathcal{E}$, $\bar{J}_0$, $\bar{J}_2$, $g$), where $g$ is the Landé g-factor, such that the parameters can be determined by making comparison with experiments. We use the energy cutoff as the energy unit, and set $\Delta = 0.1$ in the calculations. It is known that the result has only little dependence on $\Lambda$. Here we use $\Lambda = 2$ such that a chain of length 14 for each channel is sufficient to obtain in-gap states with convergence. In the DMRG calculations, the bond dimension is set to ~ 3000, leaving the truncation error less than $10^{-5}$. A proper extrapolation with the truncation error shows that the energy deviation is quite small and is of the order of $10^{-3}\Delta$. In the calculation of LDOS, the Lehmann representation in Eq.(4) is used with further taking into account the contributions from the different channels, and the delta-functions in Eq.(4) are approximated by the Gaussian function with a broadening factor $b \sim \Delta/300$ [2].

### 5. Fixing all the model parameters:

To make quantitative comparison between numerical simulation and experiments, we need to firstly determine the parameters in the model Eq. (11), (16), (17), i.e., $\mathcal{D}$, $\mathcal{E}$, $\bar{J}_0$, $\bar{J}_2$ and $g$, We determine those parameters following the steps below.

### a) Fixing anisotropy parameters $\mathcal{D}$, $\mathcal{E}$

A key feature observed in experiments is that the energy splitting of in-gap states is linearly dependent on $|B|$, and its values are different under $B_x, B_y$ and $B_z$ for the same field strength $|B|$, i.e., $\Delta E_x \neq \Delta E_y \neq \Delta E_z$. Specifically, at external fields 1.9T, $\Delta E_z$ (0.22meV) > $\Delta E_x$ (0.14meV) > $\Delta E_y$ (0.08meV), as shown by Fig.3(a) of main text. We note that the ratios between different splitting $\Delta E_x/\Delta E_z \approx 0.636$ and $\Delta E_y/\Delta E_z \approx 0.364$ are important quantities that can determine the model parameters, as we find that these ratios are barely affected by $J_0$, $J_2$, and external fields $B$, but are mainly dependent on $\mathcal{D}$, $\mathcal{E}$. Hence, $\Delta E_x/\Delta E_z \approx 0.636$ and $\Delta E_y/\Delta E_z \approx 0.364$ can be used to uniquely determine the anisotropy parameters $\mathcal{D}$, $\mathcal{E}$.

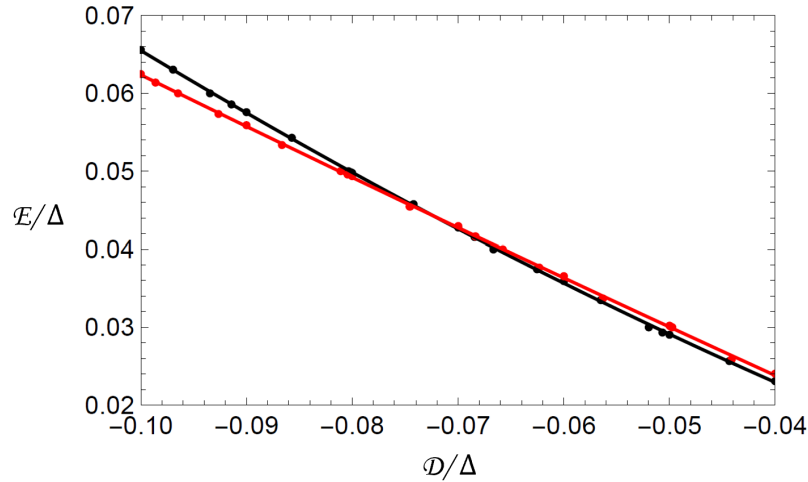

**FIG. S6.** The black (red) trajectory in the D-E plane denotes the "set of parameters" in which $\Delta E_x/\Delta E_z \approx 0.636$ ($\Delta E_y/\Delta E_z \approx 0.364$) is satisfied. Other parameters are $\bar{J}_0/\Delta$=0.95, $\bar{J}_2/\Delta$=0.65 and $g\mu_B B_i/\Delta$=0.02 where $(i = x, y, z)$.

By varying $\mathcal{D}$ and $\mathcal{E}$ in DMRG calculations, we obtain the energy of the ground state and several lowest excited in-gap states and then evaluate the splitting $\Delta E_x$, $\Delta E_y$ and $\Delta E_z$, with fixing other parameters to be $\bar{J}_0/\Delta$=0.95, $\bar{J}_2/\Delta$=0.65 and $g\mu_B B_i/\Delta$=0.02, where $(i = x, y, z)$. By imposing $\Delta E_x/\Delta E_z \approx 0.636$, we can arrive at a trajectory in the $\mathcal{D}$-$\mathcal{E}$ plane, as shown by red in Fig. S6 (i.e., on the red trajectory, $\Delta E_x/\Delta E_z \approx 0.636$ is satisfied). Then, imposing $\Delta E_y/\Delta E_z \approx 0.364$, another black trajectory is found. The crossing point then uniquely fix the parameters, $\mathcal{D}/\Delta=-0.072$, $\mathcal{E}/\Delta=0.044$, such that the splitting behaviors quantitatively agrees with the experiments.

### b) Fixing the exchange interactions $\bar{J}_0$ and $\bar{J}_2$

The second experimental feature is that there emerge two pairs of in-gap states, $\alpha\pm$, and $\beta\pm$, located around $E_{\alpha\pm} = \pm 3.1\text{meV}$ and $E_{\beta\pm}=\pm 5\text{meV}$, respectively (Fig.1(c) of main text). Since the in-gap states are symmetric in their energy location with respect to $\omega = 0$, it is sufficient to focus only on the $\alpha+$ and $\beta+$ state (we mention in passing that the $\pm$ states are asymmetric in their amplitude of LDOS. This can be well described by introducing an additional on-site scattering potential term at the impurity site that breaks the particle-hole symmetry. This is however not our main focus in this work). Using $\mathcal{D}/\Delta = -0.072$, $\mathcal{E}/\Delta = 0.044$ that are fixed in step a), we can determine the parameters $\bar{J}_0$ and $\bar{J}_2$ by using $E_{\alpha+}/\Delta = \frac{3.1\text{meV}}{7\text{meV}} = 0.443$ and $E_{\beta+}/\Delta = \frac{5.0\text{meV}}{7\text{meV}} = 0.714$ without external magnetic field, where we take $\Delta \approx 7\text{meV}$ according to the location of the inner coherent peak in Fig.1(b) of the main text. As shown by Fig. S7, the red and black trajectories in the $\bar{J}_0$-$\bar{J}_2$ plane are obtained by imposing $E_{\alpha+}/\Delta = 0.443$ and $E_{\beta+}/\Delta = 0.714$, respectively. The crossing point then uniquely fixes the exchange couplings to be $\bar{J}_0/\Delta = 0.91$ and $\bar{J}_2/\Delta = 0.64$.

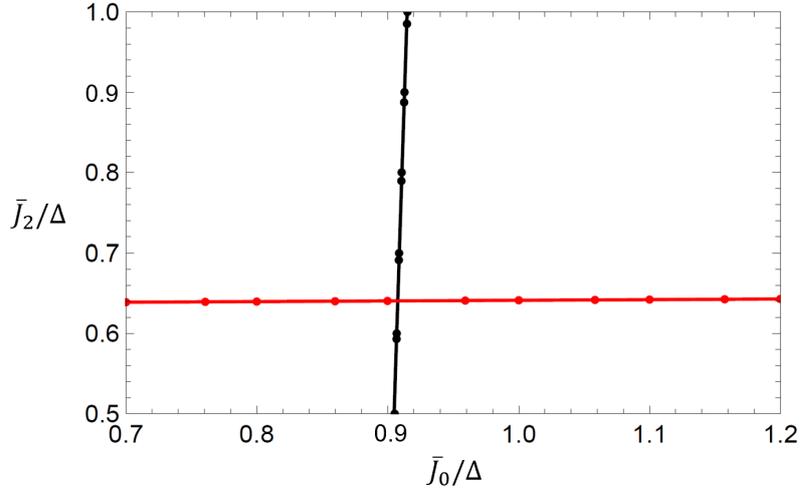

**Fig. S7.** The black (red) trajectory in the $\bar{J}_0$-$\bar{J}_2$ plane denotes the "set of parameters" in which $E_{\alpha+}/\Delta = 0.443$ ($E_{\beta+}/\Delta = 0.714$) is satisfied. Other parameters are $\mathcal{D}/\Delta = -0.072$, $\mathcal{E}/\Delta = 0.044$, and $g\mu_B B_i/\Delta=0$ where $(i = x, y, z)$.

### c) Fixing the Landé g-factor

We now extract the Landé factor by calculating the splitting of the $\alpha+$ state under varying $B_z$, as shown by Fig. S8. We fit the calculated data points (in red) by a linear function, leading to $\Delta E_z=2.84 g\mu_B B_z$. Then, after comparing to the experiments (inset to Fig. 3(a) of the main text), the $g$ factor is obtained as $g = 1.6/2.84 = 0.56$. Note that this is about 1/3 of the value estimated in the main text, which is obtained based on a S=1/2 impurity case. This is consistent with the S=2 impurity model analyzed here, because the splitting here is contributed by the splitting of an effective 3/2 spin (after screening by the $m = 0$ channel), rather than a S=1/2 state, as indicated by Fig. S5. In Fig. S8 we also calculated $\Delta E_x$ and $\Delta E_y$, as well as those for the β+ state. We can see that all of them show linear dependence on the applied magnetic fields, and the splitting magnitudes of α and β only have very small difference.

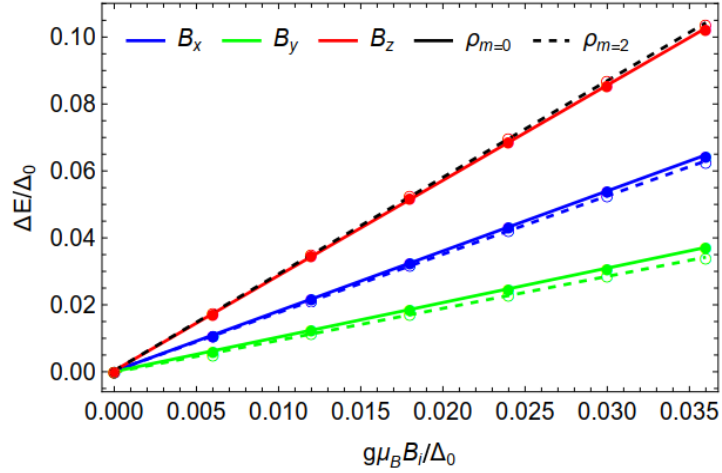

**Fig. S8.** The calculated energy splitting of the in-gap states as a function of the applied magnetic field. The red, blue green lines correspond to $\Delta E_z$ $\Delta E_x$ and $\Delta E_y$ under $B_x$ $B_y$ and $B_z$, respectively. The solid lines (dashed lines) are results for the $\alpha(\beta)$ state. Other parameters are $\mathcal{D}/\Delta = -0.072$, $\mathcal{E}/\Delta = 0.044$, $\bar{J}_0/\Delta = 0.91$ and $\bar{J}_2/\Delta = 0.64$.

## 6. Comparison between numerical simulations and experiments.

Having determined all the model parameters $\mathcal{D}$, $\mathcal{E}$, $\bar{J}_0$, $\bar{J}_2$ and $g$, now we can examine the splitting and shift behavior of in-gap states (for both the α+ and β+ state) by further varying the external fields in different directions $B_x, B_y$ and $B_z$, with an aim to make quantitative comparisons with experiments. Before making detailed comparisons with varying $B$, we firstly show in Fig. S9 the calculated LDOS (at the impurity site) for an arbitrary external field.

As shown in Fig. S9, there are two sets of split peaks, whose centers are located around $\omega/\Delta \sim 0.43$, and $\omega/\Delta \sim 0.72$, respectively, i.e., $\omega \sim 3.0$ meV and $\omega \sim 5.0$ meV ($\Delta = 0.1$ is identified to be the experimental value of 7meV), in agreement of the set of α+ and β+ states observed in experiments. The emergence of two sets of in-gap states is due to the presence of two different energy scales of the multi-channel impurity model, i.e., $\bar{J}_0$, $\bar{J}_2$. Moreover, Fig. S9 clearly shows that the splitting behavior is anisotropy, with $\Delta E_z > \Delta E_x > \Delta E_y$, in consistent with Fig. 3(b) of the main text. Note that the same splitting behavior is found not only for the α+ but also for the β+ state.

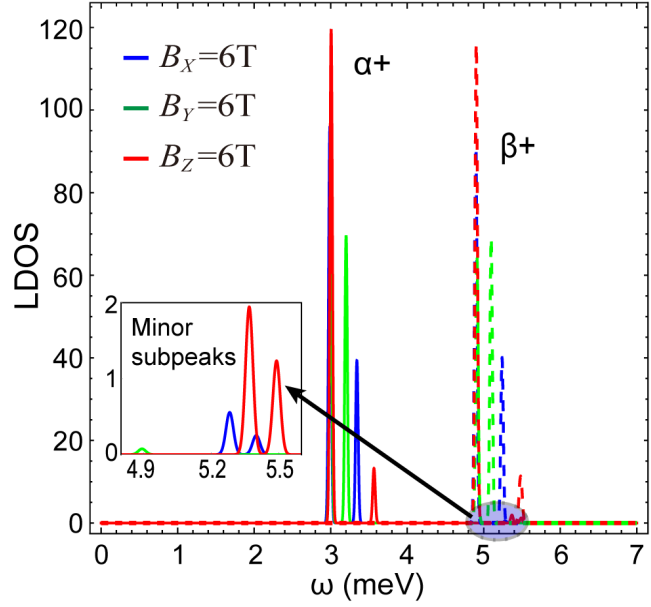

**Fig. S9.** The LDOS at the impurity site for a fixed external field. The model parameters have been fixed to $\mathcal{D}/\Delta = -0.072$, $\mathcal{E}/\Delta = 0.044$, $\bar{J}_0/\Delta = 0.91$ and $\bar{J}_2/\Delta = 0.64$. The inset shows the zoom-in data of the minor subpeaks indicated by the shaded regime. We note that the splitting of subpeaks is largest under $B_y$, which is opposite to the splitting behavior of the α+ (and β+) state. These are interesting features but their observation is yet to be experimentally confirmed and is limited by the finite resolution in realistic measurements.

We also note that there are a few "minor subpeaks" emerging in the LDOS with their centers being around $\omega/\Delta \sim 0.8$, as shown in the zoom-in data for the shaded region in Fig. S9. These minor subpeaks originate from the high-order transition process, indicated by the dashed arrows in Fig. S5. As shown in Fig. S5, under the fixed parameters, the S=3/2 states have the dominant contribution to the $|\psi_{1,2}^-\rangle$ state and little contribution to the $|\psi_{1,2}^+\rangle$ state. Therefore, the lowest excitation (which mainly describes the transition between the S=2 and S=3/2 states) is dominated by the transition process between the ground state $|G\rangle$ and $|\psi_{1,2}^-\rangle$, which has much larger amplitude than that between $|G\rangle$ and $|\psi_{1,2}^+\rangle$. Thus, the transition between $|G\rangle$ and $|\psi_{1,2}^+\rangle$ leads to much less visible LDOS peaks, i.e., the "minor subpeaks" in Fig. S9. The zoom-in data in the inset of Fig. S9 shows that the maxima of these minor subpeaks are around 1 percent of the amplitude of the main peaks. Thus, they should not be visible in realistic experiments and are neglected in the following comparisons.

In the following, we will make careful comparisons between the simulations and the experiments in terms of more details in the following three aspects.

**a) The splitting and its evolution with field $B$ for α+ and β+ state.**

We calculate the splitting of both the α+ and the β+ state along three directions, $\Delta E_x$, $\Delta E_y$, $\Delta E_z$, with varying magnetic fields. As shown in Fig. S10(a), the splitting strength is of similar values for α+ and β+ state, in consistent with the experiments (Fig. 3(a) in the main text). Besides, the splitting increases linearly with magnetic fields, which is expected and agrees with the experiments (the inset of Fig. 3(a)). We then compare the detailed splitting values between the simulation and the experiments. As shown in Fig. S10(b), the calculated splitting agrees well at the quantitative level with the experimental data extracted from Fig.3(a). We also plot in Fig. S10(c) the detailed evolution of the full LDOS with increasing $B_z$ from 0 to 8.5T. These results are in satisfying agreement with Fig. 3(a).

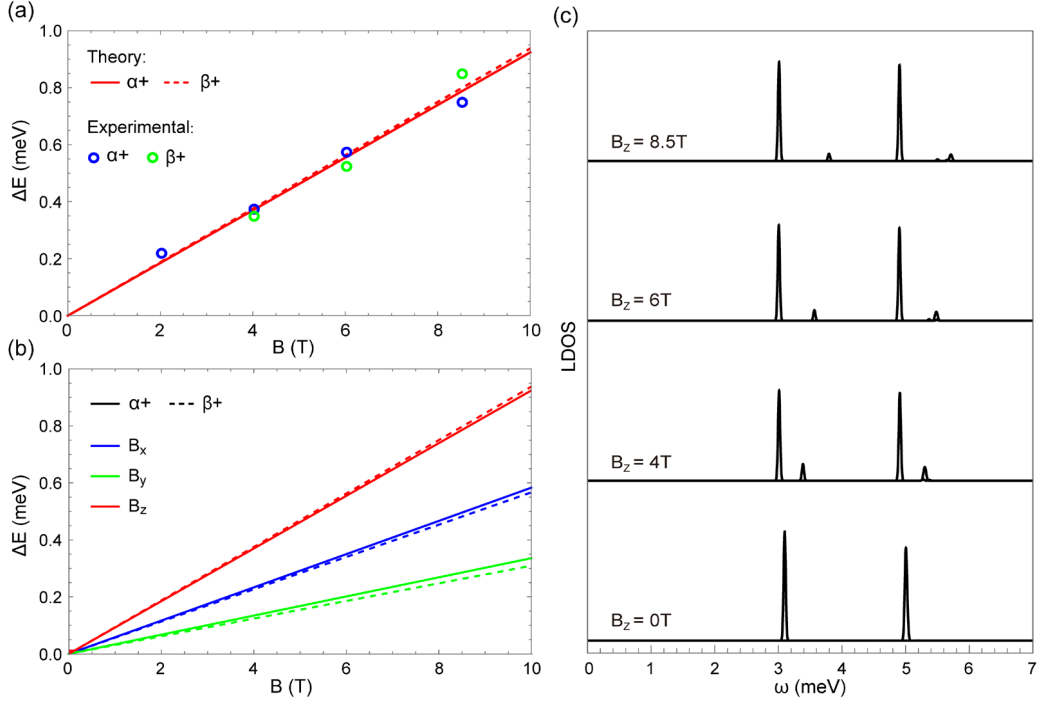

**Fig. S10.** The evolution of the splitting along x, y, z directions of the α and β state with varying external magnetic fields. (a) shows the quantitative comparison between the calculated $\Delta E_z$ and the experimental values of α+ and β+ (denoted by black data points). (b) shows the calculated full $B$ dependence of all splitting energies $\Delta E_{x,y,z}$ for both the α and β state. (c) displays the full LDOS of the in-gap states for $B_z = 0, 4, 6, 8.5$T, which is consistent with Fig.3(a) of the main text.

**b) The peak shift and its evolution with *B* for α+ and β+ state**

We also calculate the shift of the peaks under magnetic fields. Specifically, we firstly extract the center of the split peaks (the mean value of the two maxima of the LDOS peaks) and then trace its motion under increasing magnetic fields. The solid curves in Fig. S11 (a) and (b) show the energy positions of the two split peaks of the α+ and β+ state with varying $B_x$, $B_y$, $B_z$, respectively. The dotted curves are their mean values (the centers) as functions of $B_x$, $B_y$, $B_z$. It is seen the centers slightly shifts to higher energy for increasing $B_z$ or $B_y$ but such shift is less significant with comparing to the splitting. In Fig. S11(c) we plot the measured peak shift of α+ and β+ at increased $B_z$ field (with subtracting the original peak position at B=0T). One can see the peak center indeed shifts to higher energy when $B_z \lesssim 4$T, consistent with the calculation. However, at high fields ($B_z \gtrsim 6$T) the center shifts to low energy. We speculate such nonlinear shift could be due to the influence of magnetic vortex at high $B_z$. As show in Fig. S11(d), at $B_z = 8.5$T, three vortices emerge around the measured defect (no vortex is pinned on this defect). These vortices will inevitably weaken the superconductivity around the defect, which will reduce the gap size and thus lower the energy of YSR state. We note a similar nonlinear YSR peak shift at high field was also reported in Phys. Rev. Res. 4, 033182 (2022) (ref. 40). A full understanding of such behavior would rely on further measurement of local superconducting gap at increased field and considering it in the model calculation.

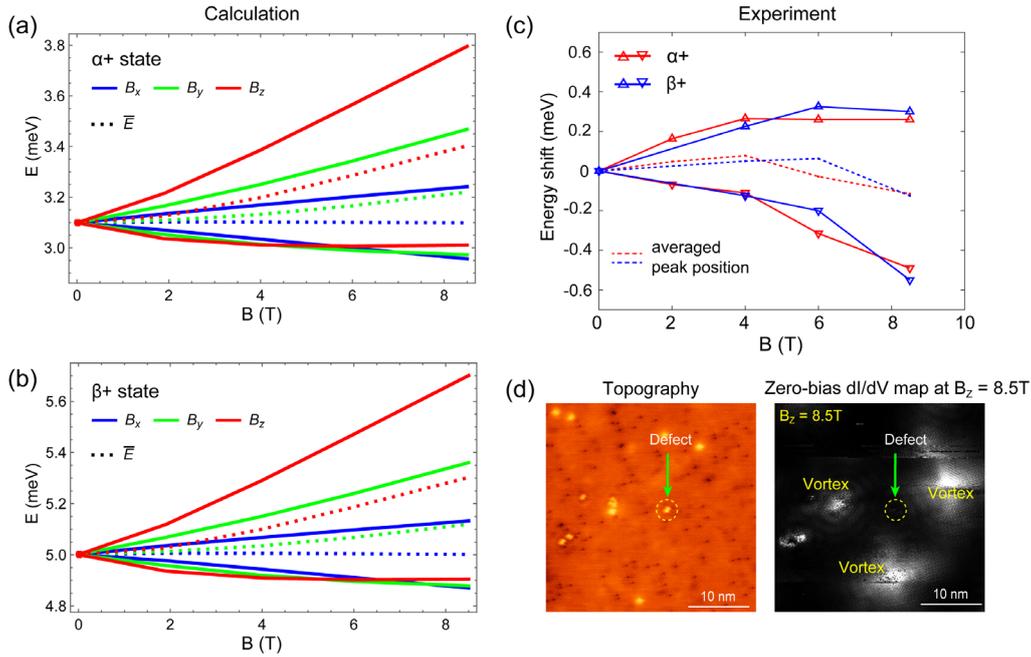

**Fig. S11**. The energy evolution of two split peaks under $B_x$, $B_y$, $B_z$ for the α+ state (a) and β+ state (b). The solid curves show the energy positions of the two split peaks, while the dashed curves denote their center positions. As shown, the shift of the peak centers under increasing $B$ is less significant than the splitting effect. (c) The measured energy shift of the two split peaks of α+ and β+ state, with respect to their original position at B=0T (d) left: STM topography showing the measured dumbbell defect; right: the corresponding zero-bias dI/dV map taken at B=8.5T. Several vortices (high DOS regions) are observed around the defect at B=8.5T.

### c) The anisotropic splitting with a fixed field strength.

Last, we fix the magnetic field to $B = 1.9$T and show the detailed LDOS in Fig. S12 The upper, middle, and lower panel are the results for $B_x, B_y, B_z = 1.9$ T, respectively. We find that the calculated LDOS display several detailed features in accordance with Fig. 3(b) of the main text, which is reproduced in Fig. S12(a): (1) the splitting $\Delta E_x$, $\Delta E_y$, $\Delta E_z$ match well with the experimental data, i.e., 0.14meV, 0.08meV, and 0.22meV, (2) the relative amplitude of the two split peaks also agrees qualitatively with that of Fig. S12(a), namely, the left peak is higher than the right peak, and the difference is most manifested under $B_z$, then $B_x$, and finally $B_y$. We note that the left peak being higher is due to the larger transition rate between |G> to $|\psi_1^-\rangle$ than to $|\psi_2^-\rangle$ in a magnetic field, as indicated by the blue arrows in Fig.S5. This is because under $B_z$, the weight of the state $S_z = -2$ is enlarged in |G>, while the weight of the state $S_z = -3/2$ is also enhanced in $|\psi_1^-\rangle$. Therefore, the transition rate from |G> to $|\psi_1^-\rangle$ becomes larger, leading to higher peak at lower energy.

These details further confirm that the proposed model, i.e., the multi-channel impurity model in superconductors with magnetic anisotropy, to be a satisfying description of our experiments. These results also advance the conventional knowledge of impurity physics in superconductors.

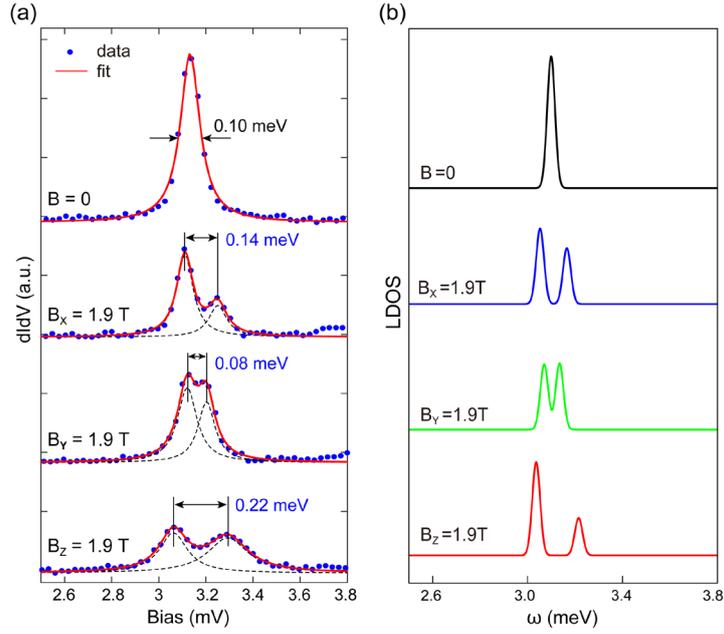

**Fig. S12**. Comparison between experiments (a) and numerical simulations (b) in terms of the detailed features of the LDOS for $B_x = 1.9$T, $B_y = 1.9$T, or $B_z = 1.9$T. The model parameters have been fixed according to the discussions above. Both the calculated splitting and the relative amplitude of the LDOS peaks are in agreement with experiments.